\documentclass[10pt,showpacs,amsmath,amssymb]{article}
\usepackage{amsfonts}
\usepackage{mathdots}
\usepackage{color}

\begin{document}

\title{Constant curvature surfaces of the supersymmetric $\mathbb{C}P^{N-1}$ sigma model}

\author{L. Delisle${}^{1,5}$, V. Hussin${}^{1,2,6}$, \.{I}. Yurdu\c{s}en${}^{3,7}$ and W. J. Zakrzewski${}^{4,8}$}

\footnotetext[1]{D\'epartement de Math\'ematiques et de
Statistique, Universit\'e de Montr\'eal, C.P. 6128,
Succ.~Centre-ville, Montr\'eal (Qu\'ebec) H3C 3J7, Canada.}
\footnotetext[2]{Centre de Recherches Math\'ematiques,
Universit\'e de Montr\'eal, C.P. 6128, Succ.~Centre-ville,
Montr\'eal (Qu\'ebec) H3C 3J7, Canada.}
\footnotetext[3]{Department of Mathematics, Hacettepe University, 06800 Beytepe, Ankara, Turkey.}
\footnotetext[4]{Department of Mathematical Sciences, University of Durham, Durham DH1 3LE, United Kingdom.}
\footnotetext[5]{email:delisle@dms.umontreal.ca}
\footnotetext[6]{email:hussin@dms.umontreal.ca}
\footnotetext[7]{email:yurdusen@hacettepe.edu.tr}
\footnotetext[8]{email:w.j.zakrzewski@durham.ac.uk}
\date{\today}

\maketitle

\begin{abstract}
Constant curvature surfaces are constructed from the finite action solutions of the supersymmetric $\mathbb{C}P^{N-1}$ sigma model. It is shown that there is a unique holomorphic solution which leads to constant curvature surfaces: the generalized Veronese curve. We give a general criterion to construct non-holomorphic solutions of the model. We extend our analysis to general supersymmetric Grassmannian models. 
\end{abstract}


\section{Introduction}

Studies of exact solutions of  integrable models are a subject of great interest in the mathematics and physics communities. In particular, the bosonic integrable $\mathbb{C}P^{N-1}$ sigma model (in 2 euclidean dimensions) has found applications in physics, biology and mathematics \cite{gross,safram,davydov,rajaraman,landolfi}. It also solves a linear spectral problem \cite{wojtekbook,grundlandy,grundland}.

This bosonic model is described by a complex vector field  $Z_b:\mathbb{R}^2 \to \mathbb{C}^{N}$ which satisfies  $\vert Z_b\vert^2=1$. Note that we have put an extra label $b$ to emphasise that we are dealing with a purely bosonic model; later we will generalise this to the supersymmetric case in which a similar field is a superfield {\it i.e.} is a function on a superspace.  The field $Z_b$ satisfies equations which describe the critical points of the action functional
 \begin{equation}
\mathcal{S}_{b}=\int_{R^2}dx_+\,dx_-\,\mathcal{L}_{b},\quad \mathcal{L}_{b}=2(\vert D_{x_+}Z_b\vert^2+\vert D_{x_-}Z_b\vert^2).\label{actionb}
\end{equation}

In this expression we use the complex variables $x_{\pm}=x\pm i y$ to parametrise $\mathbb{R}^2$ and have introduced
the covariant derivatives $D_{x_{\pm}}\Lambda=\partial_{x_{\pm}}\Lambda-\Lambda (Z_b^{\dagger}\partial_{x_{\pm}}Z_b)$.

It is easy to see that these critical points of (\ref{actionb}) are given by fields $Z_b$ which solve the Euler-Lagrange equations
\begin{equation}
D_{x_+}D_{x_-}Z_b+Z_b\vert D_{x_-}Z_b\vert^2=0. \label{EL}
\end{equation}
 
If we now restrict our attention to solutions of (\ref{EL}) of finite action $\mathcal{S}_{b}$, we have to require that $D_{x_{\pm}}Z_b\rightarrow 0$ for $\vert x\pm iy\vert\rightarrow \infty$. This compactifies $\mathbb{R}^2 \sim S^2$ via the stereographic projection. All such critical point maps are then called harmonic maps in the mathematical literature \cite{helein}. As all such maps are characterised by 
a topological number (corresponding to the values of $\pi_2(\mathbb{C}P^{N-1})$) some of these solutions have been also interpreted as static topological solitons \cite{wojtekbook,manton}
of the same model in (2+1) dimensions. 

In fact, all such harmonic maps have been known for some time
(see {\it e.g.} \cite{wojtekbook}). They can be  
obtained from the recursive application of an orthogonalization operator ${P}_{x_+}$ to a holomorphic solution $f(x_+)$ in order to obtain, in the general case, $N-2$ mixed fields ${P}_{x_+}^if$, for $i=1,2\cdots, N-2$, and a anti-holomorphic field ${P}_{x_+}^{N-1}f$. Thus, they are given by 
\begin{equation}
Z_{b,k}=\frac{{P}_{x_+}^kf}{\vert {P}_{x_+}^kf\vert},\quad k=0,1,\cdots,N-1,\label{classol}
\end{equation}
where $Z_{b,0}$ is sometimes referred to as an instanton, while $Z_{b,i}$ for $i=1,2\cdots, N-2$ as mixtures of instantons and anti-instantons and $Z_{b,N-1}$  as an anti-instanton solution. Note that we have
\begin{equation}
P_{x_+}^0f=f,\quad P_{x_+}f=\partial_{x_+}f-\frac{f^{\dagger}\partial_{x_+}f}{\vert f\vert^2}f ,\quad \quad P_{x_+}^kf=P_+(P_{x_+}^{k-1}f),\quad k=1,\cdots,N-1.
\label{Pplus}
\end{equation}

The $\mathbb{C}P^{N-1}\cong G(1,N)$ bosonic sigma model, as well as its generalization to Grassmannian  $G(M,N)$ bosonic model \cite{wojtekbook}, can be formulated also in a gauge invariant way in terms of rank $M$ orthogonal projectors \cite{wojtekbook,manton,post}. Indeed, from a generic solution $Z_b$ of the model, the rank one orthogonal projector $\mathbb{P}_b=Z_bZ_b^{\dagger}\in \mathbb{C}^{N\times N}$, satisfying $\mathbb{P}_b=\mathbb{P}_b^{\dagger}=\mathbb{P}_b^2$ and $\hbox{Tr} \ \mathbb{P}_b=1$, is a solution of the equivalent form of the Euler-Lagrange equations (\ref{EL}):
\begin{equation}
[\partial_{x_+}\partial_{x_-}\mathbb{P}_b,\mathbb{P}_b]=0.\label{ELP}
\end{equation}
This gauge invariant formulation of the $\mathbb{C}P^{N-1}$ model played an essential role in the construction of surfaces associated with solutions of the model \cite{grundlandy,grundland,grundland1,del1,del2,Hussin}.

Few years ago, the Weierstrass representation of surfaces in multidimensional spaces \cite{helein,kono,kono1,kono2} such as Lie algebras and groups has generated interest in surfaces associated with the solutions $Z_{b,k}=(\ref{classol})$ of the $\mathbb{C}P^{N-1}$ bosonic sigma model \cite{grundlandy,grundland,post,grundland1,del1,del2,Hussin,bolton}. Indeed, these solutions were shown to possess interesting geometric properties and they live in the $su(N)$ Lie algebra \cite{grundlandy,grundland}.

 To construct such surfaces, it was realized that a special operator which is a solution of a conservation law associated to the model, generated an $su(N)$ closed one-form $\alpha_b$ \cite{post,grundland1}. Indeed, from the Euler-Lagrange equation (\ref{ELP}), we get the conservation law
\begin{equation}
\partial_{x_+}\mathbf{L}_b-\partial_{x_-}\mathbf{L}_b^{\dagger}=0,\quad \mathbf{L}_b=[\partial_{x_-}\mathbb{P}_b,\mathbb{P}_b],\label{consbos}
\end{equation}
from which we can construct the $su(N)$ valued one-form
\begin{equation}
\alpha_b=\mathbf{L}_b\, dx_-+\mathbf{L}_b^{\dagger}\, dx_+,
\end{equation}
which is closed whenever $\mathbb{P}_b$ is a solution of the Euler-Lagrange equations (\ref{ELP}). From the Poincar\'e lemma \cite{warner,arnold} and the fact that $S^2$ is compact, we deduce that $\alpha_b$ is also exact and so can be used to construct a surface $\mathbf{X}_b$ from its tangent vector as
\begin{equation}
\partial_{x_+}\mathbf{X}_b=\mathbf{L}_b^{\dagger},\quad \partial_{x_-}\mathbf{X}_b=\mathbf{L}_b.
\label{derXb}
\end{equation}

The surface $\mathbf{X}_b$ naturally lives in the Lie algebra $su(N)$ for a convenient choice of the constant of integration, and from the scalar product 
\begin{equation}
(A, B)=\frac12 \hbox{Tr} (A B),\quad A,B\in su(N),\label{scalarprod}
\end{equation}
we can calculate various geometric properties of these surfaces such as the induced metric and the gaussian curvature \cite{grundland,post,grundland1,del1,del2,Hussin,bolton}. Indeed, from the first fundamental form $\mathcal{I}= (d\mathbf{X}_b,d\mathbf{X}_b)$, we find that the components of the metric tensor $g_b$ are given by
\begin{equation}
g_{b,++}=g_{b,--}^{\dagger}=\frac{1}{2}\hbox{Tr}(\mathbf{L}_b^{\dagger})^2,\quad g_{b,+-}=g_{b,-+}=\frac{1}{2}\hbox{Tr}(\mathbf{L}_b\mathbf{L}_b^{\dagger})
\end{equation}
or, in terms of the projector $\mathbb{P}_b$ by
\begin{equation}
g_{b,++}=g_{b,--}^{\dagger}=-\frac12\hbox{Tr}(\partial_{x_+}\mathbb{P}_b\partial_{x_+}\mathbb{P}_b),\quad g_{b,+-}=g_{b,-+}=\frac{1}{2}\hbox{Tr}(\partial_{x_-}\mathbb{P}_b\partial_{x_+}\mathbb{P}_b).
\label{metbos}
\end{equation}

Then, for the solutions $\mathbb{P}_b$ of (\ref{ELP}), one can show \cite{grundland,post,grundland1,del1,del2,Hussin} that the metric components $g_{b,++}=0$ and thus we deduce the explicit form of the gaussian curvature \cite{helein,gray}:
\begin{equation}
\mathcal{K}_b=-\frac{1}{g_{b,+-}}\partial_{x_+}\partial_{x_-}\ln g_{b,+-}.\label{gausscurv}
\end{equation}

Other geometric properties, such as the second fundamental form, the Willmore functional and the mean curvature have also been determined \cite{grundland,post}. In this paper we limit ourselves to the study of the gaussian curvature. This choice is motivated by the fact that two decades ago, Bolton \textit{et al.} in \cite{bolton} studied various properties of the surfaces obtained from the solutions of the $\mathbb{C}P^{N-1}$ bosonic sigma model. They were particularly interested in finding conditions for these surfaces to be of constant gaussian curvature. In \cite{bolton} they presented a complete classification of such surfaces in terms of the Veronese curve and gave also some pinching theorems. In fact they showed that, up to gauge transformations, the only solutions which lead to constant curvature surfaces are those given by $Z_{b,k}=(\ref{classol})$ for which $f$ is chosen as the Veronese curve
\begin{equation}
f(x_+)=\left(\,1\,,\,\sqrt{{N-1\choose 1}}x_+\,,\cdots,\, \sqrt{{N-1\choose r}}x_+^r\,,\cdots,\, x_+^{N-1}\,\right).\label{Veronese}
\end{equation}
For further discussions on the construction of such surfaces, we refer the reader to the Appendix where we recall how the surfaces associated to holomorphic solutions naturally lives in the Lie algebra $su(N)$ and show that they can be immersed in $\mathbb{R}^{N^2-1}$ where they describe spheres of prescribed radius.

A classification of such solutions for a general Grassmannian $G(M,N)$  bosonic sigma model \cite{wojtekbook} is not complete and has been the object of extensive research work \cite{del1,del2,Hussin,jiao,jiao1,snobl,jie}. Recently, three of us (LD, VH and WJZ) obtained some new results and gave two conjectures relating to the possible values of the constant gaussian curvature of such solutions \cite{del1}  and, recently, these conjectures have got partial support \cite{del2,peng,peng1}.

The aim of this paper is to consider the supersymmetric (SUSY) generalization of this problem for which little is known. General considerations of the SUSY $\mathbb{C}P^{N-1}$ sigma model and its solutions have been given in several papers \cite{witten,Adda,Fujii}.  A good review is given in  \cite{wojtekbook}. 
The constraints which appear in the usual formulation of this model have limited some authors to the study of the constant curvature surfaces obtained from only the holomorphic solutions of the SUSY $\mathbb{C}P^1$ model \cite{Hussin2}. In this paper, we consider the SUSY $\mathbb{C}P^{N-1}$ model and we show that, up to gauge symmetry, the holomorphic solutions corresponding to the constant curvature surfaces are related to a generalization of the Veronese curve (\ref{Veronese}). Furthermore, we also present the explicit forms of the non-holomorphic solutions and show that, for the non trivial $\mathbb{C}P^2$ model, these solutions are heavily constrained as predicted by the authors in \cite{Hussin2}. 

In Section 2, we present a description of the SUSY $\mathbb{C}P^{N-1}$ sigma model and discuss a systematic way of constructing surfaces, via the Weierstrass representation, from solutions of this model. Section 3 is devoted to the construction of special invariant solutions of the model and of the constant curvature surfaces obtained from them. Then, in Section 4, we consider the case of SUSY holomorphic solutions of the model and present a theorem that shows that the only solutions with constant curvature are in fact associated with what we call a generalized holomorphic Veronese curve. Section 5 presents a discussion of non-holomorphic solutions of the model. In Section 6, we discuss possible generalizations of our results to the SUSY $G(M,N)$ model. We finish the paper with concluding remarks and our plans for future studies.

\section{SUSY $\mathbb{C}P^{N-1}$ sigma model}

In this section, we consider the two-dimensional SUSY $\mathbb{C}P^{N-1}$ sigma model and show how to generate surfaces from its solutions following a procedure similar to the one used
 in the bosonic model.

\subsection{The model}
The SUSY $\mathbb{C}P^{N-1}$ sigma model can be constructed using a two-dimensional superspace $(x_1,x_2;\theta_1,\theta_2)$ where $(x_1,x_2)$ are local coordinates on $\mathbb{R}^2$ and $(\theta_1,\theta_2)$ are components of a Majorana spinor regarded as being real. This superspace $(x_1,x_2,\theta_1,\theta_2)$, whose bosonic part is compactified to the $2$-sphere $S^2$, will be denoted by $\tilde S^2$. For future considerations, we work in the two-dimensional complex superspace $(x_+,x_-;\theta_+,\theta_-)$, that will also be denoted by $\tilde S^2$, where
\begin{equation}
x_{\pm}=x_1\pm ix_2,\quad \theta_{\pm}=\theta_1\pm i\theta_2.
\end{equation}

The dynamical variable field of the model is a vector bosonic superfield $\Phi$ defined on the complex superspace which has the following finite Taylor expansion
\begin{equation}
\Phi(x_+,x_-;\theta_+,\theta_-)=z(x_+,x_-)+i\theta_+\chi_+(x_+,x_-)+i\theta_-\chi_-(x_+,x_-)-\theta_+\theta_-F(x_+,x_-),\label{dynvar}
\end{equation}
where $z$ and $F$ are $N$-components bosonic vectors and $\chi_+$ and $\chi_-$ are $N$-components fermionic vectors. As in the bosonic sigma model (\ref{EL}), the field $\Phi$ satisfies the nonlinear constraint 
\begin{equation}
\vert \Phi\vert^2=\Phi^{\dagger}\Phi=1,\label{nonlinearcons}
\end{equation}
where the complex conjugation of $\Phi$ is defined naturally as
\begin{equation}
\Phi^{\dagger}=z^{\dagger}+i\theta_-\chi_+^{\dagger}+i\theta_+\chi_-^{\dagger}-\theta_+\theta_-F^{\dagger}.\label{conjugation}
\end{equation}
The explicit form of the constraint (\ref{nonlinearcons}) may be found by direct calculations (see for example \cite{wojtekbook, Hussin2}).

The action is defined in this case as \cite{wojtekbook,witten,Adda,Fujii,Hussin2}
\begin{equation}
\mathcal{S}=\int_{\tilde S^2}d\theta_+d\theta_-dx_+dx_-\mathcal{L}=\int_{S^2}dx_+dx_-(\partial_{\theta_+}\partial_{\theta_-}\mathcal{L}),\label{action}
\end{equation}
where the Lagrangian density $\mathcal{L}$ is given by 
\begin{equation}
\mathcal{L}=2(\vert\check{D}_+\Phi\vert^2-\vert\check{D}_-\Phi\vert^2).\label{lagrangian}
\end{equation}

We note that the action (\ref{action}) depends only  on the $\theta_+\theta_-$ component of the Lagrangian density $\mathcal{L}$ and, as a consequence, the finiteness of the action  depends only  on this component of the Lagrangian. The supercovariant derivatives $\check{D}_{\pm}$ are defined as
\begin{equation}
\check{D}_{\pm}\Lambda=\check{\partial}_{\pm}\Lambda-\Lambda(\Phi^{\dagger}\check{\partial}_{\pm}\Phi),
\end{equation}
where the usual superderivatives 
\begin{equation}
\check{\partial}_{\pm}=-i\partial_{\theta_{\pm}}+\theta_{\pm}\partial_{x_\pm}
\end{equation}
are SUSY invariant operators. Both these derivatives anticommute with the supercharges $\check{Q}_{\pm}$ \cite{Corn} given by
\begin{equation}
\check{Q}_{\pm}=i\partial_{\theta_{\pm}}+\theta_{\pm}\partial_{x_\pm}.
\end{equation}
and all these operators satisfy the usual anticommutation rules
\begin{eqnarray}
\{\check{\partial}_{-},\check{\partial}_{+}\}&=&0,\quad \check{\partial}_{\pm}^2=-i\partial_{x_\pm}, \nonumber \\
\{\check{Q}_-,\check{Q}_+\}&=&0,\quad \check{Q}_{\pm}^2=i\partial_{x_\pm},\nonumber \\
\{\check{Q}_\pm,\check{\partial}_{\pm}\}&=&0,\quad \{\check{Q}_\pm,\check{\partial}_{\mp}\}=0.
\end{eqnarray}

Using the  principle of least action, we find that the superfield $\Phi$ satisfies the Euler-Lagrange equations given as
\begin{equation}
\check{D}_+\check{D}_-\Phi+\vert\check{D}_-\Phi\vert^2\Phi=0,\label{ELequa}
\end{equation}
together with the constraint (\ref{nonlinearcons}). In order to obtain finite action solutions of the SUSY sigma model, we have to impose the boundary conditions
\begin{equation}
\check{D}_{\pm}\Phi\rightarrow 0,\quad \vert x_\pm\vert\rightarrow \infty.
\end{equation}

\subsection{The projector formalism and surfaces constructed from the solutions of the SUSY model}

Here we discuss our construction of surfaces in the Lie algebra $su(N)$ using a procedure similar to the one described before for the bosonic model (see Introduction). First we  consider the orthogonal projector formulation of our SUSY model. From a solution $\Phi$ of the SUSY $\mathbb{C}P^{N-1}$ sigma model, we construct, as in the bosonic model \cite{wojtekbook}, the supermap $\tilde\mathbb{P}: \tilde S^2 \to \mathbb{C}^{N\times N}$ given by
\begin{equation}
\tilde{\mathbb{P}}=\Phi\Phi^{\dagger}.\label{susyproj}
\end{equation}

The supermap $\tilde{\mathbb{P}} (x_+,x_-,\theta_+,\theta_-)$ is an orthogonal projector of rank one since, from (\ref{nonlinearcons}), it satisfies the usual bosonic constraints
\begin{equation}
\tilde{\mathbb{P}}^{\dagger}=\tilde{\mathbb{P}},\quad \tilde{\mathbb{P}}^2=\tilde{\mathbb{P}},\quad  \hbox{Tr}\ \tilde{\mathbb{P}}=1.
\label{defproj}
\end{equation} 

The Euler-Lagrange equations (\ref{ELequa}) may then be rewritten as 
\begin{equation}
[\check{\partial}_+\check{\partial}_-\tilde{\mathbb{P}},\tilde{\mathbb{P}}]=0.\label{susyEL}
\end{equation}
which turns out to give us a superconservation law \cite{Hussin2}
\begin{equation}
\check{\partial}_+[\check{\partial}_-\tilde{\mathbb{P}},\tilde{\mathbb{P}}]-\check{\partial}_-[\check{\partial}_+\tilde{\mathbb{P}},\tilde{\mathbb{P}}]=0.\label{susycons}
\end{equation}
This expression may be transformed into a bosonic conservation law by applying the $\check{\partial}_+\check{\partial}_-$ operator to it. Then we get
\begin{equation}
\partial_{x_+}{\mathbf{L}}-\partial_{x_-}{\mathbf{L}}^{\dagger}=0,\quad {\mathbf{L}}=i \check{\partial}_-[\check{\partial}_-\tilde{\mathbb{P}},\tilde{\mathbb{P}}],
\label{bosoniccons}
\end{equation}
expressions which are similar to the ones obtained in the bosonic case (\ref{consbos}) but now, we have 
\begin{equation}
{\mathbf{L}}=[{\partial}_{x_-}\tilde{\mathbb{P}},\tilde{\mathbb{P}}]- 2 i (\check{\partial}_-\tilde{\mathbb{P}})^2.
\label{formL}
\end{equation}
We would like to point out that the term $(\check{\partial}_-\tilde{\mathbb{P}})^2$ is non zero since $\tilde \mathbb{P}$ is a matrix.

Next we perform the construction of the surfaces generated by solutions of the SUSY model just as this has been done in the bosonic case and explained in the introduction. The corresponding one-form $\tilde\alpha$ is still traceless even though we have an additional contribution with respect to the bosonic case since $\hbox{Tr}(AB)=-\hbox{Tr}(BA)$ for $A$ and $B$ fermionic matrices. This form satisfies $\tilde\alpha^{\dagger}=\tilde\alpha$. The metric is still given by (\ref{metbos}) where now the projector $\tilde{\mathbb{P}}$ is a superfield.
 

Note that these results are more general than the one obtained in  \cite{Hussin2}, where only the holomorphic case was studied. Here our procedure applies also to non-holomorphic immersions. 


Introducing the canonical variable $\Phi=\frac{w}{\vert w\vert}$, we can construct the rank one orthogonal projector as defined by  (\ref{susyproj})
\begin{equation}
\tilde{\mathbb{P}}=\frac{w w^\dag}{\vert w\vert^2}.
\label{pw}
\end{equation} 
We call this particular form the canonical one since it is well defined on the equivalence classes of $[w]$ in $\mathbb{C}P^{N-1}$ and it  trivially satisfies the conditions of being a rank one orthogonal projector (\ref{defproj}). In this canonical representation, the general expression for the metric $\tilde g$ (this new notation has been introduced to emphasise that our metric is now a function of $\theta_\pm$ and so is defined on superspace) is
\begin{equation}
{\tilde g_{++}}={\tilde g_{--}}^{\dagger}=-\frac{1}{\vert w\vert^2}\partial_{x_+}w^{\dagger}(\mathbb{I}-\tilde{\mathbb{P}})\partial_{x_+}w
\label{gplus}
\end{equation}
and
\begin{equation}
{\tilde g_{+-}}={\tilde g_{-+}}=\frac{1}{2\vert w\vert^2}(\partial_{x_+}w^{\dagger}(\mathbb{I}-\tilde{\mathbb{P}})\partial_{x_-}w+\partial_{x_-}w^{\dagger}(\mathbb{I}-\tilde{\mathbb{P}})\partial_{x_+}w).
\label{gplusmoins}
\end{equation}

This paper, as it will be shown later, deals with conformally parametrized surfaces in the sense that ${\tilde g_{++}}={\tilde g_{--}}=0$ and, in this case, the gaussian curvature is given as in (\ref{gausscurv}) but involving the superfields.

\section{SUSY invariant solutions and the Veronese sequence}

In this section, we construct harmonic solutions of the SUSY sigma model  by performing SUSY translations. We show that such solutions generate a finite set of constant curvature surfaces in $su(N)$ if we restrict ourself to the Veronese sequence in a proper way. This procedure is similar as the one described for the bosonic case. The major question that will be treated in the following sections is the completeness of the set of such solutions.

Let us start by assuming that our SUSY orthogonal projector $\tilde{\mathbb{P}}$ is such that it can be written as
\begin{equation}
{\tilde{\mathbb{P}}}(x_+,x_-,\theta_+,\theta_-)={\tilde{\mathbb{P}}}(y_+,y_-),
\label{invp}
\end{equation}
where $y_+$ and $y_-$ are SUSY translated variables defined by
\begin{equation}
y_+=x_++i\theta_+\frac{\xi_1(x_+)}{\sqrt{N-1}}, \quad y_-=y_+^{\dagger}.
\end{equation}

In this case, the superconservation equation (\ref{susycons}) becomes equivalent to 
\begin{equation}
\left(\frac{\vert\xi_1\vert^2}{N-1}-\theta_+\frac{\xi_1^{\dagger}}{\sqrt{N-1}}+\theta_-\frac{\xi_1}{\sqrt{N-1}}-\theta_+\theta_-\right)[\partial_{y_+}\partial_{y_-}\tilde{\mathbb{P}},\tilde{\mathbb{P}}]=0,\label{invequa}
\end{equation}
where we have used the convention that $\vert \xi_1\vert^2=\xi_1^\dagger \xi_1$.

Finding solutions $\tilde{\mathbb{P}}=(\ref{invp})$ which satisfy (\ref{invequa}) in this case has got reduced to the ones of the bosonic case (\ref{ELP}) and we can exploit the fact that the complete set of solutions is known \cite{wojtekbook}. Indeed, starting with $w=w(y_+)$, a holomorphic superfield, the complete set of solutions is obtained by considering
$\{ \ P_{y_+}^kw, \  k=0,\cdots,N-1\}$ where the definition of $P_{y_+}$ is as in (\ref{Pplus}) with $y_+$ and $w=w(y_+)$ being now  a supervariable and a superfield respectively.

Let us now note that this set can be used to construct $N$ projectors defined as
\begin{equation}
\tilde{\mathbb{P}}_k(y_+,y_-)=\frac{P_{y_+}^kw(P_{y_+}^kw)^{\dagger}}{\vert P_{y_+}^kw\vert^2},\quad k=0,1,\cdots, N-1\label{invsol},
\end{equation}
where for $k=0$ we have a holomorphic solution and for $k=1,\cdots, N-2$ non-holomorphic (or mixed) solutions  and for $k=N-1$ an anti-holomorphic solution. Of course, each of these orthogonal projectors solves the equation
\begin{equation}
[\partial_{y_+}\partial_{y_-}\tilde{\mathbb{P}}_k(y_+,y_-),\tilde{\mathbb{P}}_k(y_+,y_-)]=0,\quad k=0,1,\cdots,N-1.
\end{equation}

Note that the holomorphic superfield $w=w(y_+)$ can be expanded as the SUSY invariant holomorphic solution
\begin{equation}
w(y_+)=u(x_+)+i\theta_+\frac{\xi_1(x_+)}{\sqrt{N-1}}{\partial_{x_+} u(x_+)}.
\label{holoinv}
\end{equation}
In consequence, the SUSY invariant non-holomorphic solutions can be expanded as ($k=1,\cdots, N-1$)
\begin{equation}
P_{y_+}^kw=\left(1+i\theta_+\frac{\xi_1(x_+)}{\sqrt{N-1}}\partial_{x_+}+i\theta_-\frac{\xi_1^{\dagger}(x_+)}{\sqrt{N-1}}\partial_{x_-}-\theta_+\theta_-\frac{\vert\xi_1(x_+)\vert^2}{N-1}\partial_{x_+}\partial_{x_-}\right){P}_{x_+}^ku.
\label{nholoinv}
\end{equation}

This shows that the purely bosonic part is given by ${P}_{x_+}^ku(x_+)$ and  the fermionic parts are functions of an arbitrary function $\xi_1(x_+)$ and of the usual derivatives of the superfield ${P}_{x_+}^ku(x_+)$.

\subsection{Constant curvature surfaces of the Veronese type}

Based on the results of Section 2, we now compute the metric and the gaussian curvature of the SUSY invariant solutions we have just obtained.

Using the property that
\begin{equation}
\partial_{y_+}\tilde{\mathbb{P}}_k=\left(1+i\theta_+\frac{\partial_{x_+}\xi_1}{\sqrt{N-1}}\right)\left(\frac{P_{y_+}^{k+1}w(P_{y_+}^kw)^{\dagger}}{\vert P_{y_+}^kw\vert^2}-\frac{P_{y_+}^kw(P_{y_+}^{k-1}w)^{\dagger}}{\vert P_{y_+}^{k-1}w\vert^2}\right),
\end{equation}
the metric components (\ref{metbos}), in the SUSY case, become 
\begin{equation}
\tilde{g}_{++}=\tilde{g}_{--}=0,\quad \tilde{g}_{+-}=\tilde{g}_{-+}=\frac{A}{2} \left(\frac{\vert P_{y_+}^{k+1}w\vert^2}{\vert P_{y_+}^kw\vert^2}+\frac{\vert P_{y_+}^kw\vert^2}{\vert P_{y_+}^{k-1}w\vert^2}\right),
\end{equation}
where $A$ is a bosonic superfield of the form
\begin{equation}
A(x_+,x_-,\theta_+,\theta_-)=1+i\theta_+\frac{\partial_{x_+}\xi_1}{\sqrt{N-1}}+i\theta_-\frac{\partial_{x_-}\xi_1^{\dagger}}{\sqrt{N-1}}-\theta_+\theta_-\frac{\vert \partial_{x_+}\xi_1\vert^2}{N-1}.
\end{equation}

This last result shows that the surfaces obtained by this procedure are still conformally parametrized and, thus, that the expression for the Gaussian curvature is given as in (\ref{gausscurv}). Next we compute this curvature for our surfaces. We find that
\begin{equation}
\ln (A)=\frac{i}{\sqrt{N-1}}(\theta_+ \partial_{x_+}\xi_1+\theta_- \partial_{x_-}\xi_1^{\dagger}),
\end{equation}
so that $\partial_{x_+}\partial_{x_-}\ln(A)=0$. In this case, the curvature is similar to the one of the purely bosonic model and takes the following explicit form:
\begin{equation}
\tilde{\mathcal{K}}_k=-2\left(\frac{\vert P_{y_+}^{k+1}w\vert^2}{\vert P_+{y_+}^kw\vert^2}+\frac{\vert P_{y_+}^kw\vert^2}{\vert P_{y_+}^{k-1}w\vert^2}\right)^{-1}\partial_{y_+}\partial_{y_-}\ln\left(\frac{\vert P_{y_+}^{k+1}w\vert^2}{\vert P_{y_+}^kw\vert^2}+\frac{\vert P_{y_+}^kw\vert^2}{\vert P_{y_+}^{k-1}w\vert^2}\right),
\end{equation}
for $k=0,1,\cdots, N-1$.

This expression was shown to be constant when $w$ was taken as the Veronese curve \cite{grundlandy,grundland,post,grundland1,del1,del2,Hussin,bolton}, \textit{i.e.} for $w(y_+)=f(y_+)$ where $f$  is the Veronese curve as in (\ref{Veronese}). Note also that the SUSY holomorphic solution coincides with (\ref{holoinv}) when $u(x_+)$ is chosen as the purely bosonic Veronese curve (\ref{Veronese}). The SUSY non-holomorphic solutions are explicitly written as (\ref{nholoinv}) with the same $u(x_+)$. 

We get the usual value \cite{del2} of the gaussian curvature:
\begin{equation}
\tilde\mathcal{K}_k=\frac{4}{N-1+2k(N-1-k)},\quad k=0,1,\cdots,N-1.
\end{equation}

In the bosonic case, we know that the surfaces constructed this way are the only solutions with constant curvature. In the next sections, we will investigate whether this is also the case in the SUSY model.

\section{SUSY holomorphic solutions with constant curvature }

In this section, we present a theorem which states that the only SUSY holomorphic solutions with constant curvature are the SUSY invariant ones given by (\ref{holoinv}). 

Let us start our discussion with the construction of the surfaces generated by a general holomorphic solution of the SUSY $\mathbb{C}P^{N-1}$ sigma model . Using the gauge invariance of the model \cite{wojtekbook,witten}, any such solution is given by (\ref{pw})      
where 
\begin{equation}
w(x_+,\theta_+)=\left(\,1\,,\,W_1(x_+,\theta_+)\,,\cdots,\, W_{N-1}(x_+,\theta_+)\,\right)^T.\label{wform}
\end{equation}
As usual, we can perform the following expansion:
\begin{equation}
w(x_+,\theta_+)=u(x_+)+i\theta_+\xi(x_+),\label{projhol}
\end{equation}
and we see that  $u$ is a purely bosonic $N$-column vector and $\xi$ is a fermionic $N$-column vector such that:
\begin{equation}
u(x_+)=\left(\,1\,,\,u_1(x_+)\,,\cdots,\,u_{N-1}(x_+)\,\right)^T, \ \xi(x_+)=\left(\,0\,,\,\xi_1(x_+)\,,\cdots,\,\xi_{N-1}(x_+)\,\right)^T.\label{wform1}
\end{equation}

From (\ref{bosoniccons}), we see  \cite{Hussin2} that $\mathbf{L}= -\partial_{x_-} \tilde\mathbb{P}$  with $\tilde\mathbb{P}=(\ref{pw})$ and, from (\ref{derXb}) and the fact that $\tilde\mathbf{X}\in su(N)$. Then equations (\ref{derXb}) may be easily integrated and for a convenient choice of the constant of integration, we have that the surface $\tilde\mathbf{X}$ is given by 
\begin{equation}
\tilde\mathbf{X}=\tilde\mathbb{P}-\frac{1}{N}\mathbb{I}_N.
\label{surfacehol}
\end{equation}
As in the bosonic case, it satisfies the properties discussed in the Appendix, even though we are dealing with superfields. 

From the expressions (\ref{gplus}) and (\ref{gplusmoins}), we see that the metric components 
take the form
\begin{equation}
\tilde g_{++}=\tilde g_{--}=0,\quad \tilde g_{+-}=\tilde g_{-+}=\frac{\vert{P}_{x_+}w\vert^2}{2\vert w\vert^2},\label{metrichol}
\end{equation}
which shows that general holomorphic immersions lead to conformally parametrized surfaces \cite{gray}. 

Using  properties of the operator ${P}_{x_+}$ given in \cite{wojtekbook,del1,del2,Hussin} and defined as in (\ref{Pplus}), we calculate the explicit form of the gaussian curvature (\ref{gausscurv}) and we obtain 
\begin{equation}
\tilde\mathcal{K}=4-2\frac{\vert w\vert^2\vert{P}_{x_+}^2w\vert^2}{\vert{P}_{x_+}w\vert^4}.\label{kahol}
\end{equation}
In \cite{Hussin2}, it was shown that, for the SUSY $\mathbb{C}P^1$ model, surfaces associated with holomorphic solutions are always of constant curvature $\mathcal{K}=4$. This result can be easily reproduced here using (\ref {kahol}), since ${P}_{x_+}^2w$ is always zero in the $\mathbb{C}P^1$ model. 

We now state our general result on the uniqueness of these holomorphic solutions.

\noindent\textbf{Theorem}: The general form of the holomorphic solution of the SUSY $\mathbb{C}P^{N-1}$ sigma model leading to constant curvature surfaces is the solution given in (\ref{holoinv}) with $u(x_+)$ being the bosonic Veronese curve defined in (\ref{Veronese}). We have named this curve the Generalized SUSY Veronese (GSV) curve.

\noindent\textbf{Proof:}
 
First we perform the general expansion of the metric component $\tilde g_{+-}$ and of the curvature $\tilde \mathcal{K}$ in terms of the components of the field $w$. Using (\ref{projhol}), we get 
\begin{equation}
|w|^2=a_0+i\theta_+a_1+i\theta_-a_2-\theta_+\theta_-a_3,
\end{equation}
with
\begin{equation}
a_0=\vert u\vert^2,\quad a_1=a_2^{\dagger}=u^{\dagger}\xi,\quad a_3=\vert\xi\vert^2.
\label{aa}
\end{equation}
Thus, from (\ref{metrichol}), we obtain 
\begin{equation}
\tilde g_{+-}=g_0+i\theta_+g_1+i\theta_-g_2-\theta_+\theta_-g_3,
\end{equation}
with \begin{equation}
g_0=\frac12\partial_{x_+}\partial_{x_-}\ln a_0,\quad g_1=g_2^{\dagger}=\frac12\partial_{x_+}\partial_{x_-}\left(\frac{a_1}{a_0}\right),\quad g_3=\frac12\partial_{x_+}\partial_{x_-}\left(\frac{a_0a_3-a_2a_1}{a_0^2}\right).\label{methol}
\end{equation}


Since the metric components $\tilde g_{++}=\tilde g_{--}=0$, we make use of expression (\ref{gausscurv}) to get
\begin{equation}
\tilde\mathcal{K}=\mathcal{K}_0+i\theta_+\mathcal{K}_1+i\theta_-\mathcal{K}_2-\theta_+\theta_-\mathcal{K}_3,
\end{equation}
with
\begin{equation}
\mathcal{K}_0=-\frac{1}{g_0}\partial_{x_+}\partial_{x_-}\ln g_0,\quad \mathcal{K}_1=\mathcal{K}_2^{\dagger}=-\frac{1}{g_0}\partial_{x_+}\partial_{x_-}\left(\frac{g_1}{g_0}\right)-\frac{g_1}{g_0}\mathcal{K}_0
\end{equation}
and
\begin{equation}
\mathcal{K}_3=-\frac{1}{g_0}\partial_{x_+}\partial_{x_-}\left(\frac{g_0g_3-g_2g_1}{g_0^2}\right)-\frac{g_3}{g_0}\mathcal{K}_0-\frac{g_2}{g_0}\mathcal{K}_1+\frac{g_1}{g_0}\mathcal{K}_2.
\end{equation}

To achieve constant curvature surfaces, we first need $\mathcal{K}_0$ to be constant. From the bosonic sigma model \cite{del1,del2,Hussin,bolton}, we know that $u=u(x_+)$ must be the Veronese curve given as in (\ref{Veronese}). We thus get, as in the classical case,
\begin{equation}
a_0=(1+\vert x\vert^2)^{N-1},\quad g_0=\frac{(N-1)}{2(1+\vert x\vert^2)^2},\quad \mathcal{K}_0=\frac{4}{N-1}.
\label{constantzero}
\end{equation}

To complete our analysis we need to determine the possible expression for $\xi(x_+)$ in (\ref{wform1}) that leads to $\mathcal{K}_1=\mathcal{K}_2=\mathcal{K}_3=0$. We will show that $\mathcal{K}_1=0$ requires that $\xi$ has the form as stated in the theorem, \textit{i.e.} $w$ is of the GSV form. This, in turn, will imply that $\mathcal{K}_2=\mathcal{K}_3=0$ since we know that  (\ref{holoinv}) is a solution of our problem from section 3.1.

Using the expressions (\ref{constantzero}) and the form of $g_1$ as in (\ref{methol}), we see that the constraint $\mathcal{K}_1=0$ reduces to 
\begin{equation}
\partial_{x_+}^2\partial_{x_-}^2 h(x_+,x_-)=0,\quad  h(x_+,x_-)=(1+\vert x\vert^2)^{2-N} a_1=(1+\vert x\vert^2)^{2-N} u^{\dagger}\xi.
\label{famousequ}
\end{equation}

In order to solve this problem, we express the superfield $\xi$ as a linear combination of the derivatives of the bosonic Veronese sequence $u$:
\begin{equation}
\xi(x_+)=\sum_{i=0}^{N-1}\phi_i(x_+)\partial_{x_+}^i u,\label{xicombination}
\end{equation}
where $\phi_i(x_+)$ are fermionic functions of $x_+$ for $i=0,1,\cdots,N-1$. This linear combination is equivalent to the matrix equation
\begin{equation}
\xi=A \phi,\quad A=\left(\,u\,,\,\partial_{x_+}u\,,\cdots,\,\partial_{x_+}^{N-1}u\,\right)\label{Amatrix}
\end{equation}
and $\phi=(0,\phi_1,\cdots,\phi_{N-1})^T$. By construction, the matrix $A$ is triangular and we get 
\begin{equation}
\det A=\prod_{i=0}^{N-1}i!\sqrt{{N-1\choose i}}=((N-1)!)^{N/2}.
\end{equation}
The system (\ref{Amatrix}) is thus invertible and, from the fact that $\det A$ is constant, the vector field $\phi$ is a function of $x_+$ as desired and is of the polynomial form. This implies that
\begin{equation}
\xi(x_+)=\frac{\xi_1(x_+)}{\sqrt{N-1}}{\partial_{x_+} u(x_+)} \iff \phi_{1}(x_+)=\frac{\xi_1(x_+)}{\sqrt{N-1}},\ \phi_2=\dots=\phi_{N-1}=0.
\end{equation}

Going back to  $h(x_+,x_-)$ given by (\ref{famousequ}), we note that
\begin{eqnarray}
h(x_+,x_-)&=&(1+\vert x\vert^2)^{2-N} \sum_{i=1}^{N-1}\phi_i u^{\dagger}\partial_{x_+}^i u
=(1+\vert x\vert^2)^{2-N} \sum_{i=1}^{N-1}\phi_i\partial_{x_+}^i a_0\nonumber\\
&=&(N-1)! \sum_{i=1}^{N-1}\frac{1}{(N-i-1)!}\frac{x_-^i \phi_i(x_+)}{(1+\vert x\vert^2)^{i-1}}.
\label{ka}
\end{eqnarray}
Since $h(x_+,x_-)$ satisfies (\ref{famousequ}), we get, by standard integration, the general solution in terms of four arbitrary functions $h_1(x_-),\ h_2(x_+),\ h_3(x_-),\ h_4(x_+)$ as
\begin{equation}
h(x_+,x_-)=h_1(x_-)x_++h_2(x_+)x_-+ h_3(x_-)+ h_4(x_+).
\label{kagen}
\end{equation}
We see that the maximal power of $(1+\vert x\vert^2)$ in the denominator of $h(x_+,x_-)$ in (\ref{ka}) is $(N-2)$ and so we can  rewrite (\ref{kagen}) as
\begin{eqnarray}
&&(N-1)! \sum_{i=1}^{N-1}\frac{(1+\vert x\vert^2)^{N-i-1}}{(N-i-1)!}x_-^i \phi_i(x_+)\nonumber\\
&=&(1+\vert x\vert^2)^{N-2}(h_1(x_-)x_++h_2(x_+)x_-+ h_3(x_-)+ h_4(x_+)).
\label{newequ}
\end{eqnarray}
In order to fix the arbitrary functions, we evaluate the expression (\ref{newequ}) at $x_+=0$ and $x_-=0$. At $x_-=0$, we get the constraint
\begin{equation}
h_4(x_+)=-h_1(0)x_+- h_3(0)
\end{equation}
and, at $x_+=0$, we have
\begin{equation}
h_3(x_-)=-h_2(0)x_--h_4(0)+h(0,x_-)=-h_2(0)x_--h_4(0)+(N-1)!\sum_{i=1}^{N-1}\frac{x_-^i\phi_i(0)}{(N-i-1)!}.
\end{equation}
Next we put these expressions of $h_4(x_+)$ and $h_3(x_-)$ into (\ref{newequ}) and we get
\begin{eqnarray}
&&(1+\vert x\vert^2)^{N-2}(N-1)(\phi_1(x_+)-\phi_1(0))x_- \nonumber\\
&+&(N-1)! \sum_{i=2}^{N-1}\frac{x_-^i}{(N-i-1)!} \left((1+\vert x\vert^2)^{N-i-1}\phi_i(x_+)-(1+\vert x\vert^2)^{N-2}\phi_i(0)\right)\nonumber\\
&=&(1+\vert x\vert^2)^{N-2}\left((h_1(x_-)-h_1(0))x_++(h_2(x_+)-h_2(0))x_-\right).
\label{newequb}
\end{eqnarray}
Note that in this last expression, we have also used the fact that $h_3(0)+h_4(0)=0$. 

Now it is easy to show that the left hand side of this expression is a polynomial in $x_-$ of degree $(2N-3)$. We thus have
\begin{equation}
h_1(x_-)-h_1(0)=\sum_{i=1}^{N-1} a_i x_-^i,
\end{equation}
where the coefficients $a_i$ are left to be determined. From this last result and by comparing the coefficients of $x_-$, we deduce that
\begin{equation}
h_2(x_+)-h_2(0)=-a_1 x_++(N-1)(\phi_1(x_+)-\phi_1(0)).
\end{equation}
Taking into account all theses constraints, the expression (\ref{newequb}) becomes
\begin{eqnarray}
&&(N-1)! \sum_{i=2}^{N-1}\frac{x_-^i}{(N-i-1)!} \left((1+\vert x\vert^2)^{N-i-1}\phi_i(x_+)-(1+\vert x\vert^2)^{N-2}\phi_i(0)\right)\nonumber\\
&=&(1+\vert x\vert^2)^{N-2}\sum_{i=2}^{N-1} a_i x_-^ix_+.
\end{eqnarray}
Dividing this last expression by $x_-^2$, we obtain
\begin{equation}
\sum_{i=2}^{N-1}\frac{x_-^{i-2}(1+\vert x\vert^2)^{N-i-1}}{(N-1-i)!}\phi_i(x_+)-\sum_{i=2}^{N-1}\frac{x_-^{i-2}(1+\vert x\vert^2)^{N-2}}{(N-1-i)!}(\phi_i(0) +b_i x_+)=0
\label{newequbb}
\end{equation}
with $b_i=\frac{(N-i-1)!}{(N-1)!} a_i$. This last expression is a polynomial in the variable $x_-$ of maximal degree $(2N-5)$. Furthermore, we can observe that the first summation is a polynomial expression of $x_-$ of maximal degree $(N-3)$. Let us then consider the $x_-^j$'s terms for $j> N-3$ of the above polynomial. These terms are all contained in the second summation and we get the coefficients of $x_-^j$ for $j=N-2,\cdots, 2N-5$ which are proportional to
\begin{equation}
\phi_{j-N+4}(0)+b_{j-N+4}x_+.
\end{equation}
These expressions must vanish for (\ref{newequbb}) to be satisfied. It shows that $\phi_i(0)=0$ and $b_i=a_i=0$ for $i=2,\cdots, N-1$. Hence the expression (\ref{newequbb}) reduces to
\begin{equation}
\sum_{i=2}^{N-1}\frac{x_-^{i-2}}{(N-1-i)!} \left((1+\vert x\vert^2)^{N-i-1}\phi_i(x_+)\right)=0.
\end{equation}
From this expression, it is easy to see that $\phi_i(x_+)=0$ for $i=2,\cdots, N-1$. This proves our theorem. 
Interestingly, we see also that the arbitrary functions are given as
\begin{equation}
h_1(x_-)=h_1(0)+a_1x_-,\quad h_2(x_+)=h_2(0)-a_1x_++(N-1)(\phi_1(x_+)-\phi_1(0))
\end{equation}
and 
\begin{equation}
h_3(x_-)=-h_2(0)x_-+h_3(0)+(N-1)x_-\phi_1(0),\quad h_4(x_+)=-h_1(0)x_+-h_3(0).
\end{equation}

Let us conclude this section by giving the expression for the metric. It generalizes the result obtained for the SUSY $\mathbb{C}P^1$ model and is of the form:
\begin{eqnarray}
\tilde g_{+-}&=&\frac{N-1}{2}\bigg(\frac{1}{(1+\vert x\vert^2)^2}+i\theta_+\partial_{x_+}\left(\frac{\phi_1(x_+)}{(1+\vert x\vert^2)^2}\right)+i\theta_-\partial_{x_-}\left(\frac{(\phi_1(x_+))^\dagger}{(1+\vert x\vert^2)^2}\right)\nonumber \\
& &-\theta_+\theta_-\partial_{x_+}\partial_{x_-}\left(\frac{\vert\phi_1\vert^2}{(1+\vert x\vert^2)^2}\right)\bigg).
\end{eqnarray}
We see that the metric has fermionic contributions as expected, but they are all total derivatives and vanish after integration over $x_+$ and $x_-$.

\section{A set of non-holomorphic solutions of SUSY $\mathbb{C}P^{N-1}$ with constant curvature}

 An avenue for getting non-holomorphic solutions of the SUSY $\mathbb{C}P^{N-1}$ model involves reproducing the steps of a proof for the bosonic sigma model given in \cite{wojtekbook}. Interestingly, it relates an holomorphic solution of the $G(m,N)$ model to a non-holomorphic solution of the $\mathbb{C}P^{N-1}$ model. More precisely, let us consider the holomorphic projector $\mathbf{P}_{b}^{(m-1)}=\sum_{j=0}^{m-1}\mathbb{P}_{b,j}$, solution of the $G(m,N)$ model, and the non-holomorphic projector $\mathbb{P}_{b,m}$ of the $\mathbb{C}P^{N-1}$ model which is orthogonal $\mathbf{P}_{b}^{(m-1)}$. It was thus shown (\cite{wojtekbook}) that $\mathbb{P}_{b,m}$ solves the Euler-Lagrange equations. This was based on the following property:
\begin{equation}
0=({\partial}_{x_-}(\mathbf{P}_{b}^{(m-1)}+\mathbb{P}_{b,m}))(\mathbf{P}_{b}^{(m-1)}+\mathbb{P}_{b,m})=({\partial}_{x_-}\mathbb{P}_{b,m})\mathbb{P}_{b,m}+
{\partial}_{x_-}\mathbf{P}_{b}^{(m-1)},
\end{equation}
since $({\partial}_{x_-}\mathbf{P}_{b}^{(m-1)})\mathbf{P}_{b}^{(m-1)}=({\partial}_{x_-}\mathbb{P}_{b,m})\mathbf{P}_{b}^{(m-1)}=0$ and $({\partial}_{x_-}\mathbf{P}_{b}^{(m-1)})\mathbb{P}_{b,m}={\partial}_{x_-}\mathbf{P}_{b}^{(m-1)}$, which leads to
\begin{equation}
[{\partial}_{x_+}{\partial}_{x_-}\mathbb{P}_{b,m},\mathbb{P}_{b,m}]=0.
\end{equation}

Now, we consider SUSY projectors. We construct a set of $N$ orthogonal projectors
\begin{equation}
\check{\mathbb{P}}_i=\frac{w_i\otimes w_i^{\dagger}}{\vert w_i\vert^2},\quad i=0,1,\cdots,N-1,
\end{equation}
where the $w_i$'s are constructed recursively as 
\begin{equation}
w_0=\varphi_0(x_+,\theta_+),\quad w_i=(\mathbb{I}-\check{\mathbf{P}}^{(i-1)})\varphi_i,\quad \check{\mathbf{P}}^{(i-1)}=\sum_{j=0}^{i-1}\check{\mathbb{P}}_j.
\end{equation}
The $\varphi_i$'s are holomorphic supervectors for $i=0,1,\cdots,N-1$. By construction the $w_i$'s are mutually orthogonal and we choose the $\varphi_i$'s in such a way that $w_i\neq 0$ for all $i$. Using a different approach then the one of MacFarlane's \cite{macfarlane}, we can show:

\noindent\textbf{Proposition 1:} The projector $\check{\mathbf{P}}^{(m)}$ is an holomorphic solution of the $G(m+1,N)$ model for $m=0,1,\cdots,N-2$ without any restrictions on the supervectors $\varphi_i$, in the sense that it satisfies the following relation
\begin{equation}
\check{\mathbf{P}}^{(m)}\check{\partial}_-\check{\mathbf{P}}^{(m)}=\check{\partial}_-\check{\mathbf{P}}^{(m)}.
\end{equation}

\noindent\textbf{Proof:}  We proceed by induction. For $\check{\mathbf{P}}^{(0)}=\check{\mathbb{P}}_0$, we have that
\begin{equation}
\check{\partial}_-\check{\mathbf{P}}^{(0)}=\frac{\varphi_0\otimes (\check{P}_+\varphi_0)^{\dagger}}{\vert \varphi_0\vert^2}\quad \Longrightarrow\quad \check{\mathbf{P}}^{(0)}\check{\partial}_-\check{\mathbf{P}}^{(0)}=\check{\partial}_-\check{\mathbf{P}}^{(0)}
\end{equation}
and, as a consequence, $\check{\mathbf{P}}^{(0)}$ is an holomorphic solution without any restriction on the holomorphic supervector $\varphi_0$.  Now let us suppose that $\check{\mathbf{P}}^{(m-1)}$ is an holomorphic solution of $G(m,N)$ without any restrictions on the $\varphi$'s and let us show that $\check{\mathbf{P}}^{(m)}=\check{\mathbf{P}}^{(m-1)}+\check{\mathbb{P}}_m$ is an holomorphic solution of $G(m+1,N)$ again without any restrictions. Let us calculate the expression
\begin{equation}
\check{\mathbf{P}}^{(m)}\check{\partial}_-\check{\mathbf{P}}^{(m)}-\check{\partial}_-\check{\mathbf{P}}^{(m)}
\end{equation}
and show that it is zero. We have
\begin{eqnarray*}
\check{\mathbf{P}}^{(m)}\check{\partial}_-\check{\mathbf{P}}^{(m)}-\check{\partial}_-\check{\mathbf{P}}^{(m)}&=&(\check{\mathbf{P}}^{(m-1)}+\check{\mathbb{P}}_m)\check{\partial}_-(\check{\mathbf{P}}^{(m-1)}+\check{\mathbb{P}}_m)-\check{\partial}_-(\check{\mathbf{P}}^{(m-1)}+\check{\mathbb{P}}_m)\\
&=&\check{\mathbf{P}}^{(m-1)}\check{\partial}_-\check{\mathbb{P}}_m+\check{\mathbb{P}}_m\check{\partial}_-\check{\mathbf{P}}^{(m-1)}+\check{\mathbb{P}}_m\check{\partial}_-\check{\mathbb{P}}_m-\check{\partial}_-\check{\mathbb{P}}_m\\
&=&(\check{\mathbf{P}}^{(m)}-\mathbb{I})\check{\partial}_-\check{\mathbb{P}}_m\\
&= &(\check{\mathbf{P}}^{(m)}-\mathbb{I})\frac{(\check{\partial}_-w_m)\otimes w_m^{\dagger}}{\vert w_m\vert^2}.
\end{eqnarray*}
We thus see that $\check{\mathbf{P}}^{(m)}\check{\partial}_-\check{\mathbf{P}}^{(m)}-\check{\partial}_-\check{\mathbf{P}}^{(m)}=0$ if and only if
\begin{equation}
(\check{\mathbf{P}}^{(m)}-\mathbb{I})\check{\partial}_-w_m=0,\quad w_m=(\mathbb{I}-\check{\mathbf{P}}^{(m-1)})\varphi_m.
\end{equation}
We thus have
\begin{equation}
\check{\partial}_-w_m=-(\check{\partial}_-\check{\mathbf{P}}^{(m-1)})\varphi_m=-(\check{\mathbf{P}}^{(m-1)}\check{\partial}_-\check{\mathbf{P}}^{(m-1)})\varphi_m.
\end{equation}
The result is thus proven since we have that
\begin{equation}
(\check{\mathbf{P}}^{(m)}-\mathbb{I})\check{\mathbf{P}}^{(m-1)}=(\check{\mathbf{P}}^{(m-1)}+\check{\mathbb{P}}_m-\mathbb{I})\check{\mathbf{P}}^{(m-1)}=((\check{\mathbf{P}}^{(m-1)})^2-\check{\mathbb{P}}_m\check{\mathbf{P}}^{(m-1)}-\check{\mathbf{P}}^{(m-1)})=0,
\end{equation}
using $(\check{\mathbf{P}}^{(m-1)})^2=\check{\mathbf{P}}^{(m-1)}$ and $\check{\mathbb{P}}_m\check{\mathbf{P}}^{(m-1)}=0$. This concludes the proof.

We, now, use this relation and a similar approach as in \cite{wojtekbook}, to find a criterion for the projector $\check{\mathbb{P}}_m$ to be a solution of the SUSY $\mathbb{C}P^{N-1}$ model. Indeed, let us write $\check{\mathbf{P}}^{(m)}=\check{\mathbf{P}}^{(m-1)}+\check{\mathbb{P}}_m$ and, thus, using the elements of the proof above we show directly that 
\begin{equation}
\check{\mathbf{P}}^{(m-1)}\check{\partial}_-\check{\mathbb{P}}_m=\check{\partial}_-\check{\mathbb{P}}_m-\check{\mathbb{P}}_m\check{\partial}_-\check{\mathbb{P}}_m.\label{relahol}
\end{equation}
We, then, consider the complex conjugate of equation (\ref{relahol}):
\begin{equation}
(\check{\partial}_+\check{\mathbb{P}}_m)\check{\mathbf{P}}^{(m-1)}=\check{\partial}_+\check{\mathbb{P}}_m-(\check{\partial}_+\check{\mathbb{P}}_m)\check{\mathbb{P}}_m.\label{relaholoconj}
\end{equation}
Finally, taking the $\check{\partial}_+$ of equation (\ref{relahol}) and summing it to the $\check{\partial}_-$ of equation (\ref{relaholoconj}), we get
\begin{equation}
[\check{\partial}_+\check{\partial}_-\check{\mathbb{P}}_m,\check{\mathbb{P}}_m]=\check{\partial}_+(\check{\mathbf{P}}^{(m-1)}\check{\partial}_-\check{\mathbb{P}}_m)+\check{\partial}_-(\check{\mathbf{P}}^{(m-1)}\check{\partial}_-\check{\mathbb{P}}_m)^{\dagger}.\label{relacondPm}
\end{equation}

We can thus formulate the following proposition:\\
\textbf{Proposition 2:} If we impose the condition $\check{\mathbf{P}}^{(m-1)}\check{\partial}_-\check{\mathbb{P}}_m=-(\check{\partial}_-\check{\mathbf{P}}^{(m-1)})\check{\mathbb{P}}_m=\check{\partial}_-\mathbb{A}_m$ with $\mathbb{A}_m=\mathbb{A}_m^{\dagger}$, then the projector $\check{\mathbb{P}}_m$ is a non-holomorphic solution of the SUSY $\mathbb{C}P^{N-1}$ model.

This proposition follows directly from equation (\ref{relacondPm}) and note that the first equality of the condition is a consequence of the orthogonality relation $\check{\mathbf{P}}^{(m-1)}\check{\mathbb{P}}_m=0$. The condition of proposition 2 is realized in the bosonic model, where we have $\mathbf{P}_{b}^{(m-1)}\partial_{x_-}\mathbb{P}_{b,m}=-{\partial}_{x_-}\mathbf{P}_{b}^{(m-1)}$. One property of SUSY models is that it must reduce to the bosonic model in the fermionic limit. Using this approach, we set $\mathbb{A}_m=-\check{\mathbf{P}}^{(m-1)}$ which, using proposition 2, leads to the condition
\begin{equation}
(\check{\partial}_-\check{\mathbf{P}}^{(m-1)})\check{\mathbb{P}}_m=\check{\partial}_-\check{\mathbf{P}}^{(m-1)}.\label{condisol}
\end{equation}

Let us analyse the case $m=1$. In this case, the condition (\ref{condisol}) reduces to
\begin{equation}
(\mathbb{I}-\check{\mathbb{P}}_1)\check{P}_+\varphi_0=0\quad \iff\quad (\mathbb{I}-\check{\mathbb{P}}_0-\check{\mathbb{P}}_1)\check{\partial}_+\varphi_0=0.\label{constraintP1}
\end{equation}

The general solution to this equation (or constraint) was discussed in \cite{Din}. Indeed, the authors modified the fermionic fields to being c-numbers (commuting numbers) by taking $\varphi_1=\epsilon_+\check{\partial}_+\varphi_0$ which implies that $w_1=\epsilon_+\check{P}_+\varphi_0$. Making the canonical choice $\vert \epsilon_+\vert=1$, the authors have successfully solved this constraint formally but questions were raised concerning the physical interpretation of this procedure. 

We suggest another way of solving this constraint. Indeed, in this paper, we are interested in conformally parametrized surfaces of constant curvature. Our ultimate goal is to generalize the results of Bolton and \textit{al.} \cite{bolton} which gave a complete classification of such surfaces for the bosonic $\mathbb{C}P^{N-1}$ model. This classification has turned out to be a hard mathematical problem for more general Grassmannians where only partial results have been obtained \cite{Hussin,del1,del2,jiao,jiao1,snobl,jie,peng,peng1}. Here, we generalize the result of Bolton to the SUSY extension of the $\mathbb{C}P^{N-1}$ model.

Using the fact that the metric components of the surfaces (\ref{metbos}) are the same as in the bosonic case, we construct the following SUSY projectors
\begin{equation}
\tilde{\mathbb{P}}_i=\frac{P_{x_+}^iw(P_{x_+}^iw)^{\dagger}}{\vert P_{x_+}^iw\vert^2},\quad i=0,1,\cdots,N-1,
\end{equation}
which ensures their conformal parametrization. Note that the projectors $\tilde{\mathbb{P}}_i$ are obtained from the projectors $\check{\mathbb{P}}_i$ by setting $\varphi_i=\partial_{x_+}^iw$ for $i=0,1,\cdots,N-1$ and $w=w(x_+,\theta_+)$. 

In this case, the constraint (\ref{constraintP1}) becomes
\begin{equation}
(\mathbb{I}-\tilde{\mathbb{P}}_0-\tilde{\mathbb{P}}_1)\check{\partial}_+w=0,
\end{equation}
and the completeness relation
\begin{equation}
\sum_{i=0}^{N-1}\tilde{\mathbb{P}}_i=\mathbb{I}\label{completude}
\end{equation}
gives the equivalent constraint:
\begin{equation}
(\tilde{\mathbb{P}}_2+\cdots+\tilde{\mathbb{P}}_{N-1})\check{\partial}_+ w=0
\end{equation}
or, explicitly:
\begin{equation}
\left(\sum_{k=2}^{N-1} \frac{(P_{x_+}^{k} w)(P_{x_+}^{k} w)^\dagger}{\vert P_{x_+}^k w\vert^2}\right)\check{\partial}_+ w=0.
\end{equation}
Multiplying from the left by $(P_{x_+}^jw)^{\dagger}$ for $j=2,\cdots,N-1$, and due to the orthogonality of the set $\{ P_{x_+}^{j} w, j=2,\dots,N-1\}$, we get the following set of equations
\begin{equation}
(P_{x_+}^{j} w)^\dagger\check{\partial}_+ w=0,\quad  j=2,\dots,N-1.
\end{equation}
We can further reduce this set of equations by observing that $\check{\partial}_+w=-i\partial_{\theta_+}w+\theta_+\partial_{x_+}w$ and that $(P_{x_+}^jw)^{\dagger}\partial_{x_+}w=0$ for $j=2,\cdots,N-1$. This shows that the above set is equivalent to
\begin{equation}
(P_{x_+}^jw)^{\dagger}\xi=0, \quad \xi=-i\partial_{\theta_+}w, \quad j=2,\cdots,N-1,
\label{constraintxi}
\end{equation}
where $w$ is given by (\ref{projhol}) and (\ref{wform1}).
The key ingredient to solve this set of equations is to assume, as in the holomorphic case, that $\xi$ takes the form (\ref{xicombination}) where $\phi_0(x_+)=0$ by gauge invariance. 

Then we note that the set of equations (\ref{constraintxi}) for $\theta_+=\theta_-=0$, reduces to
\begin{equation}
\sum_{i\geq j}^{N-1} \left((P_{x_+}^ju)^{\dagger} (\partial_{x_+}^i u)\right) \phi_i(x_+)=0,\quad j=2,\cdots,N-1,
\label{constraintxiu}
\end{equation}
since it can be easily seen from the orthogonality of the set $\{P_{x_+}^{j} u, j=2,\dots,N-1\}$ that we have 
$(P_{x_+}^{j}u)^{\dagger} (\partial_{x_+}^i u)=0$ for $i<j$. For $j=N-1$, we get $\phi_{N-1}\vert P_{x_+}^{N-1}u\vert^2=0$, which leads to $\phi_{N-1}=0$. Next, in a similar way, we look at equation (\ref{constraintxiu}) for $j=N-2$ and we obtain
$\phi_{N-2}\vert P_{x_+}^{N-2}u\vert^2=0$, which shows that $\phi_{N-2}=0$. We then repeat this procedure for other values of $j$ and find that $\phi_{j}=0$ for $j=2,\dots,N-1$ and so that
\begin{equation}
\xi=\phi_1(x_+)\partial_{x_+} u.
\end{equation}
We already know that, with such an expression for $\xi$, $\tilde{\mathbb{P}}_1$ is a solution of the SUSY Euler-Lagrange equations.
It corresponds to the SUSY invariant solution described in Section 3. Moreover, we recover, as well the complete set of solutions (\ref{nholoinv}) with constant curvature.

\section{Non-holomorphic solutions of the $G(M,N)$ model}

In this section, we use the completeness relation (\ref{completude}) to deduce some non-holomorphic solutions of more general Grassmannians $G(M,N)$.

Before doing so, we would like to point out that the completeness relations gives the complete set of solutions in the particular case of $\mathbb{C}P^1$. Indeed, we have two projectors $\tilde{\mathbb{P}}_0$ and $\tilde{\mathbb{P}}_1$ corresponding, respectively, to holomorphic and anti-holomorphic solutions. For $\tilde{\mathbb{P}}_0$, using gauge invariance as displayed in (\ref{wform}), we get the holomorphic solution
\begin{equation}
\tilde{\mathbb{P}}_0=\frac{1}{1+\vert W\vert^2}\left(\begin{array}{cc}
1&W^{\dagger}\\
W&\vert W\vert^2
\end{array}\right).
\end{equation}
Using the completeness property (\ref{completude}), we get that the projector $\tilde{\mathbb{P}}_1$ is given by
\begin{equation}
\tilde{\mathbb{P}}_1=\mathbb{I}-\tilde{\mathbb{P}}_0=\frac{1}{1+\vert W\vert^2}\left(\begin{array}{cc}
\vert W\vert^2&-W^{\dagger}\\
-W&1
\end{array}\right).
\end{equation}
Thus another way of showing that the SUSY Euler-Lagrange are satisfied is to use the relation
\begin{equation}
[\check{\partial}_+\check{\partial}_-\tilde{\mathbb{P}}_1,\tilde{\mathbb{P}}_1]=[\check{\partial}_+\check{\partial}_-\tilde{\mathbb{P}}_0,\tilde{\mathbb{P}}_0]=0.
\end{equation}

Moreover, the completeness relation (\ref{completude}) gives a one-to-one correspondence between the solutions of the $G(2,N)$ and $G(N-2,N)$ models. Indeed, starting with a solution of $G(2,N)$, say $\tilde{\mathbb{P}}_{i_1}+\tilde{\mathbb{P}}_{i_2}$, we get from the completeness relation the solution $\sum_{j\neq 1,2}\tilde{\mathbb{P}}_{i_j}$ of $G(N-2,N)$ as
\begin{equation}
\sum_{j\neq 1,2}\tilde{\mathbb{P}}_{i_j}=\mathbb{I}-(\tilde{\mathbb{P}}_{i_1}+\tilde{\mathbb{P}}_{i_2}),
\end{equation}
in the following sense
\begin{equation}
\left[\check{\partial}_+\check{\partial}_-\left(\sum_{j\neq 1,2}\tilde{\mathbb{P}}_{i_j}\right),\sum_{j\neq 1,2}\tilde{\mathbb{P}}_{i_j}\right]=[\check{\partial}_+\check{\partial}_-(\tilde{\mathbb{P}}_{i_1}+\tilde{\mathbb{P}}_{i_2}),\tilde{\mathbb{P}}_{i_1}+\tilde{\mathbb{P}}_{i_2}].\label{rela}
\end{equation}

The strategy here is to use the fact that $\tilde{\mathbb{P}}_0$ and $\tilde{\mathbb{P}}_{N-1}$ correspond, respectively, to a holomorphic and anti-holomorphic solutions of the model and then trivially solve the SUSY Euler-Lagrange equations (\ref{susyEL}). 

Using the duality property, we then ask what are the conditions on $w$ such that $\tilde{\mathbb{P}}_0+\tilde{\mathbb{P}}_{N-1}$ is a solution of the $G(2,N)$ model knowing that $\tilde{\mathbb{P}}_0$ and $\tilde{\mathbb{P}}_{N-1}$ are trivial solutions of the $G(1,N)$ model. Consequently, this procedure will also give the non-holomorphic solution $\sum_{i=1}^{N-2}\tilde{\mathbb{P}}_i$ of the $G(N-2,N)$ model. 

\noindent\textbf{Theorem:} If we take $w(x_+,\theta_+)=u(x_+)+i\theta_+\epsilon_+v(x_+)$, the projector $\tilde\mathbb{P}_0+\tilde\mathbb{P}_{N-1}$ is a non-holomorphic solution of the SUSY $G(2,N)$ sigma model where $v(x_+)$ takes the form $v(x_+)=\sum_{i=1}^{N-2}a_i(x_+)\partial_{x_+}^iu(x_+)$. Furthermore, with the same constraints, the projector $\tilde\mathbb{P}_1+\cdots+\tilde\mathbb{P}_{N-2}$ is a non-holomorphic solution of the SUSY $G(N-2,N)$ model, where by construction
\begin{equation}
\tilde\mathbb{P}_i=\frac{P_{x_+}^iw(P_{x_+}^iw)^{\dagger}}{\vert P_{x_+}^iw\vert^2},\quad i=0,1,\cdots,N-1.
\end{equation}

\textbf{Proof:} 
Let us consider the following parametrization of the orthogonal projectors $\tilde{\mathbb{P}}_0$ and $\tilde{\mathbb{P}}_{N-1}$:
\begin{equation}
\tilde\mathbb{P}_0=\frac{ww^{\dagger}}{\vert w\vert^2},\quad \tilde\mathbb{P}_{N-1}=\frac{\alpha\alpha^{\dagger}}{\vert \alpha\vert^2},\quad\alpha=w^*\wedge \partial_{x_-}w^*\wedge\cdots\wedge \partial_{x_-}^{N-2}w^*\propto P_{x_+}^{N-1}w,
\label{holantihol}
\end{equation}
where the fields $w$ and $\alpha$ are such that $\check{\partial}_-w=0$, $\check{\partial}_+\alpha=0$ and $\alpha^{\dagger}w=0$. 
This implies that we have
\begin{eqnarray}
&&[\check{\partial}_+\check{\partial}_-(\tilde{\mathbb{P}}_0+\tilde{\mathbb{P}}_{N-1}),(\tilde{\mathbb{P}}_0+\tilde{\mathbb{P}}_{N-1})]=
[\check{\partial}_+\check{\partial}_-\tilde{\mathbb{P}}_0,\tilde{\mathbb{P}}_{N-1}]+[\check{\partial}_+\check{\partial}_-\tilde{\mathbb{P}}_{N-1},\tilde{\mathbb{P}}_0]\nonumber\\
&=&\frac{1}{\vert\alpha\vert^2\vert w\vert^2}(\left[(\check{P}_-\alpha)^{\dagger}w\right] (\check{P}_-\alpha) w^{\dagger}+\left[w^{\dagger}(\check{P}_-\alpha)\right] w(\check{P}_-\alpha)^{\dagger}-\left[(\check{P}_+w)^{\dagger}\alpha\right] (\check{P}_+w)\alpha^{\dagger}\nonumber\\
&-&\left[\alpha^{\dagger}(\check{P}_+w)\right] \alpha (\check{P}_+w)^{\dagger}),\label{g2nequa}
\end{eqnarray}
where the operators $\check{P}_+$ and $\check{P}_-$ are defined as
\begin{equation}
\check{P}_+w=\check{\partial}_+w-\frac{w^{\dagger}\check{\partial}_+w}{\vert w \vert^2}w,\quad \check{P}_-\alpha=\check{\partial}_-\alpha-\frac{\alpha^{\dagger}\check{\partial}_-\alpha}{\vert \alpha\vert^2}\alpha.
\end{equation}

Using (\ref{g2nequa}), we find a constraint on the vector fields $w$ and $\alpha$ of the form
\begin{equation}
0=\alpha^{\dagger}[\check{\partial}_+\check{\partial}_-(\tilde{\mathbb{P}}_0+\tilde{\mathbb{P}}_{N-1}),(\tilde{\mathbb{P}}_0+\tilde{\mathbb{P}}_{N-1})]\check{P}_+w\quad \Longleftrightarrow\quad \vert \check{P}_+w\vert^2\alpha^{\dagger}\check{\partial}_+w=0.\label{constgen}
\end{equation}
The idea is to solve this constraint and show that it forces $w$ to have the GSV form given in (40). First, we look at the expression $\alpha^{\dagger}\check{\partial}_+w$. By considering the canonical basis of $\mathbb{R}^N$ given as $\{e_1,e_2,\cdots,e_N\}$ which satisfies $e_j^{T}e_i=\delta_{ij}$, we see that
\begin{equation}
\alpha^{\dagger}\check{\partial}_+w=\left|\begin{array}{ccc}
\xi_1&\cdots&\xi_{N-1}\\
\partial_{x_+}W_1&\cdots&\partial_{x_+}W_{N-1}\\
\ldots&\ddots&\ldots\\
\partial_{x_+}^{N-2}W_1&\cdots&\partial_{x_+}^{N-2}W_{N-1}
\end{array}\right|.\label{constraintnonhol}
\end{equation}
In this formulation, we have used the fact that the vector superfield $w$ may be written using (\ref{wform}-\ref{wform1}). We may thus impose that $\xi(x_+)=\epsilon_+v(x_+)$ and we then find that
\begin{equation}
\alpha^{\dagger}\check{\partial}_+w=\epsilon_+\left|\begin{array}{ccc}
v_1&\cdots&v_{N-1}\\
\partial_{x_+}u_1&\cdots&\partial_{x_+}u_{N-1}\\
\ldots&\ddots&\ldots\\
\partial_{x_+}^{N-2}u_1&\cdots&\partial_{x_+}^{N-2}u_{N-1}
\end{array}\right|.
\end{equation}
This assumption is based on the fact that we impose that the fields $u$ and $\xi$ are elements of a real Grassmann algebra spanned by $\{1,\epsilon_+\}$ where $\epsilon_+$ is a fermionic generator satisfying $\epsilon_+^2=0$. Moreover, we can also easily show that
\begin{eqnarray}
\vert \check{P}_+w\vert^2\propto \vert u\vert^2\vert\xi\vert^2-\vert u^{\dagger}\xi\vert^2-\theta_+\vert u\vert^2\xi^{\dagger}P_{x_+}u+\theta_-\vert u\vert^2(P_{x_+}u)^{\dagger}\xi-\theta_+\theta_-\vert u\vert^2\vert P_{x_+}u\vert^2
\end{eqnarray}
and hence the constraint (\ref{constgen}) reduces to
\begin{equation}
(-\theta_+\vert u\vert^2\xi^{\dagger}P_{x_+}u-\theta_+\theta_-\vert u\vert^2\vert P_{x_+}u\vert^2)\alpha^{\dagger}\check{\partial}_+w=0.
\end{equation}
Finally, we use the fact that $\vert u\vert^2\vert P_{x_+}u\vert^2$ is invertible to note that this constraint is equivalent to $\alpha^{\dagger}\check{\partial}_+w=0$. So in the case where $\xi=\epsilon_+v$, we find that
\begin{equation}
v(x_+)=\sum_{i=1}^{N-2}a_i(x_+)\partial_{x_+}^iu
\end{equation}
for $a_i$'s arbitrary bosonic functions of $x_+$. 

It remains to demonstrate that the Euler-Lagrange equations (\ref{g2nequa}) are satisfied. This follows from the fact that
\begin{equation}
0=\check{\partial}_+(\alpha^{\dagger}w)=(\check{\partial}_+\alpha^{\dagger})w+\alpha^{\dagger}\check{\partial}_+w=(\check{\partial}_-\alpha)^{\dagger}w,
\end{equation}
which implies $w^{\dagger}\check{P}_-\alpha=\alpha^{\dagger}\check{P}_+w=0$ and returning to (\ref{g2nequa}), we see that the Euler-Lagrange equations are satisfied. We thus find the form of $w$ given in (\ref{projhol}) to be
\begin{equation}
w(x_+,\theta_+)=u(x_+)+i\theta_+\epsilon_+\sum_{i=1}^{N-2}a_i(x_+)\partial_{x_+}^i u(x_+),
\end{equation}
which is different from the GSV form (\ref{holoinv}).

We do not claim that the above result is complete. Indeed, in obtaining such a solution, we have imposed two constraints: the first one was to suppose that $\xi_i=\epsilon_+v_i$ and the second was in the choice of the anti-holomorphic vector field $\alpha$. We are pretty sure that the second constraint is correct since, as in the bosonic model, this parametrization implies that our surfaces are conformal. The first condition is more restrictive since it restricts our vector fields $u$ and $\xi$ to a specific Grassmann algebra \cite{Corn}. 

In the particular case of $N=3$, it is interesting to see that the non-holomorphic projector $\tilde{\mathbb{P}}_0+\tilde{\mathbb{P}}_2$ of $G(2,3)$  will generate the missing projector $\tilde{\mathbb{P}}_1$ of $G(1,3)$ and, thus, give us a complete classification of this specific model where we recover the constraint (\ref{holoinv}) on $w$.



\section{Conclusion and outlook}

In this paper, we have presented a construction of surfaces from the solutions of the SUSY $\mathbb{C}P^{N-1}$ sigma model. Indeed, using a gauge-invariant formulation of the model in terms of rank one orthogonal projectors, we have constructed a closed one-form from which we have deduced the tangent vectors to the surfaces. We have thus showed that these surfaces naturally live in the Lie algebra $su(N)$ and deduced some of their geometrical properties such as the metric and their gaussian curvature.

We considered first the holomorphic solutions of the model since they are the simplest solutions of the Euler-Lagrange equations. For these solutions, we have shown that they induce constant curvature solutions if the corresponding projectors are parametrized by a vector superfield $w$ of the GSV form: a generalization of the Veronese curve. This theorem is important, it showed the existence and the uniqueness of such a curve but it has given us a path to find the mixed solutions of our model. 

Indeed, we used a generalization of the procedure presented in \cite{wojtekbook,macfarlane} for constructing non-holomorphic solutions. We have thus proposed a pinching theorem on the constraints that one has to impose in order to obtain such solutions. These constraints were to suppose that the vector fields components are element of a two-dimensional complex Grassmann algebra and that our surfaces be conformally parametrized.

The preceding results have been extended to some propositions for the solutions of the SUSY $G(M,N)$ model.

This paper is only a beginning and a lot of work has still to be done. We have obtained some non-holomorphic solutions which are the SUSY invariant ones. One project should consist of proving the completeness of such solutions. Also the solutions obtained in Section 5 are the results of some imposed constraints. One could ask if we could relax these constraints by considering the components of the superfield $w$ to be elements of a general complex Grassmann algebra. Furthermore, we have imposed specific forms for our surfaces to be conformally parametrized. Could one find more general solutions?

\section*{Appendix: Isomorphism between $\mathbb{R}^{N^2-1}$ and $su(N)$}

In section 2, we have defined a surface $\mathbf{X}\in su(N)$ using its tangent vector fields (\ref{derXb}). In this appendix, we recall how we can associate to this surface a surface in $\mathbb{R}^{N^2-1}$ using a natural isomorphism between $su(N)$ and $\mathbb{R}^{N^2-1}$.
Let us write
\begin{equation}
\mathbf{X}=\sum_{i=1}^{N^2-1}a_i(x_+,x_-)K_i,\label{sunlinear}
\end{equation}
where $K_i$ is a basis of the Lie algebra $su(N)$. Using the scalar product (\ref{scalarprod}) on $su(N)$, we construct an orthonormal basis $\{K_{ij}^{\mathbb{R}},K_{ij}^{\mathbb{C}},K_i\}$ of the Lie algebra $su(N)$ as
\begin{equation}
K_{ij}^{\mathbb{R}}=E_{ij}+E_{ij},\quad K_{ij}^{\mathbb{C}}=i(E_{ij}-E_{ij}), \quad i>j, i,j=1,\dots,N,
\end{equation}
and
\begin{equation}
K_i=\frac{\sqrt{2}}{\sqrt{i(i+1)}}\left(\sum_{j=1}^i E_{jj}-i E_{(i+1)(i+1)}\right), \quad i=1,\cdots, N-1,
\end{equation}
where $E_{ij}$ are canonical $N\times N$ matrices  defined as $(E_{ij})_{kl}=\delta_{ik}\delta_{jl}$. One can show that the set $\{K_{ij}^{\mathbb{R}},K_{ij}^{\mathbb{C}},K_i\}$ forms an orthonormal basis of $su(N)$. This basis will then be used to generate a vector in $\mathbb{R}^{N^2-1}$ from the $a_i$'s defined in (\ref{sunlinear}). 

Let us present a convenient example: suppose that the surface $\mathbf{X}$ is associated to an holomorphic solution of our model. We have shown in (\ref{surfacehol}), that the surface $\mathbf{X}$ takes the explicit form 
\begin{equation}
\mathbf{X}=\mathbb{P}-\frac{1}{N}\mathbb{I}_N,
\end{equation}
where $\mathbb{I}_N$ is the $N\times N$ identity matrix and $\mathbb{P}\in \mathbb{C}^{N\times N}$ is a rank one orthogonal projector holomorphic solution of the model. The surface then can be written, using the orthonormal basis, as
\begin{equation}
\mathbf{X}=\mathbb{P}-\frac{1}{N}\mathbb{I}_N=\sum_{i=1}^{N-1}a_i K_i+\sum_{i>j}(a_{ij}^{\mathbb{R}}K_{ij}^{\mathbb{R}}+a_{ij}^{\mathbb{C}}K_{ij}^{\mathbb{C}}), 
 \label{representation}
\end{equation}
with
\begin{equation}
a_i=(\mathbb{P},K_i),\quad a_{ij}^{\mathbb{R}}=(\mathbb{P},K_{ij}^{\mathbb{R}}),\quad a_{ij}^{\mathbb{C}}=(\mathbb{P},K_{ij}^{\mathbb{C}}).
\end{equation}
The coefficients $a_{ij}^{\mathbb{R}}$ and $a_{ij}^{\mathbb{C}}$ are explicitly given as
\begin{equation}
a_{ij}^{\mathbb{R}}=\frac12(\mathbb{P}_{ij}+\bar{\mathbb{P}}_{ij}),\quad a_{ij}^{\mathbb{C}}=\frac{i}{2}(\bar{\mathbb{P}}_{ij}-\mathbb{P}_{ij}),\quad i>j, i,j=1,\dots,N
\label{coordij}
\end{equation}
and
\begin{equation}
a_i=\frac{1}{\sqrt{2i(i+1)}}\left(\sum_{j=1}^{i}\mathbb{P}_{jj}-i\,\mathbb{P}_{(i+1),(i+1)}\right), \quad i=1,\cdots, N-1.
\label{coordi}
\end{equation}

Using the representation of $\mathbf{X}$ given in (\ref{representation}), we can show that
\begin{equation}
\| \mathbf{X}\|^2=(\mathbf{X},\mathbf{X})=\sum_{i=1}^{N-1}a_i^2+\sum_{i>j}((a_{ij}^{\mathbb{R}})^2+(a_{ij}^{\mathbb{C}})^2)=\frac12\left(1-\frac{1}{N}\right),
\end{equation}
which shows that the surface represented by the coordinates  (\ref{coordij}) and (\ref{coordi}) is a sphere in $\mathbb{R}^{N^2-1}$ of radius $\sqrt{\frac12\left(1-\frac{1}{N}\right)}$ centred at the origin.
In \cite{Hussin}, the authors have obtained a similar result for the bosonic model using the constraint on the orthogonal projector $\mathbb{P}^2=\mathbb{P}$. In our approach, we have constructed the components of our surfaces using a geometric approach instead of an algebraic one.

It could be instructive to explicit an example. 

In particular, for $N=2$, the projector $\mathbb{P}$ takes the form
\begin{equation}
\mathbb{P}=\frac{1}{1+\vert W\vert^2}\left(\begin{array}{cc}
1&W^{\dagger}\\
W&\vert W\vert^2
\end{array}\right).
\end{equation}
An orthonormal basis of $su(2)$ is given by the Pauli matrices
\begin{equation}
K_{21}^{\mathbb{R}}=\left(\begin{array}{cc}
0&1\\
1&0
\end{array}\right),\quad K_{21}^{\mathbb{C}}=\left(\begin{array}{cc}
0&-i\\
i&0
\end{array}\right),\quad  K_{1}=\left(\begin{array}{cc}
1&0\\
0&-1
\end{array}\right).
\end{equation}
The coordinates of the sphere in  $\mathbb{R}^3$ are given as
\begin{equation}
a_1=\frac{1-\vert W\vert^2}{2(1+\vert W\vert^2)},\quad a_{21}^{\mathbb{R}}=\frac{W+W^{\dagger}}{2(1+\vert W\vert^2)},\quad a_{21}^{\mathbb{C}}=\frac{i(W^{\dagger}-W)}{2(1+\vert W\vert^2)}
\end{equation}
and they satisfy the equation
\begin{equation}
a_1^2+(a_{21}^{\mathbb{R}})^2+(a_{21}^{\mathbb{C}})^2=\frac14.
\end{equation}

For more results on the non-holomorphic solutions in the bosonic case, see for example, \cite{grundlandy} and reference therein.
 \section*{Acknowledgements}
This work has been supported in part by research grants from Natural sciences and engineering research council of Canada (NSERC). Laurent Delisle also acknowledges a Fonds de recherche du Qu\'ebec--Nature et technologies (FQRNT) fellowship.

\end{document}